\documentclass[11pt, a4paper]{article}
\pdfoutput=1 

\usepackage[height=8.85in,width=6.75in]{geometry}

\usepackage{graphicx,rotating}     
\usepackage[bookmarksopen,colorlinks=true,linkcolor=royalblue,
citecolor=orangered,urlcolor=orangered,linktoc=all]{hyperref}

\usepackage{amsmath,amssymb}
\usepackage{slashed}
\usepackage{bm}
\usepackage{bbm}
\usepackage{cite}
\usepackage{CJKutf8}
\usepackage{comment}
\usepackage[utf8]{inputenc}
\usepackage{accents}
\usepackage[footnotesize]{caption}
\usepackage{cancel}
\usepackage{physics}
\usepackage{stmaryrd}
\usepackage[T1]{fontenc}
\usepackage{lmodern}

\usepackage{chngcntr}
\counterwithin{equation}{section}

\usepackage{xcolor}
\definecolor{darkred}{rgb}{0.7, 0., 0.}
\definecolor{orangered}{rgb}{1,0.27,0.}
\definecolor{steelblue}{rgb}{0.275,0.51, 0.706}
\definecolor{royalblue}{rgb}{0.2549,0.4118,0.8824}
\definecolor{forestgreen}{rgb}{0.13,0.55,0.13}
\definecolor{rossoferrari}{HTML}{D9073D}
\definecolor{mediumblue}{HTML}{0000CD}

\usepackage{tikz}
\usepackage{tikz-feynman}
\tikzfeynmanset{compat=1.1.0}
\pgfdeclarelayer{bg}    
\pgfsetlayers{bg,main}  
\usetikzlibrary{decorations}

\pgfdeclaredecoration{dashsoliddouble}{initial}{
  \state{initial}[width=\pgfdecoratedinputsegmentlength]{
    \pgfmathsetlengthmacro\lw{.3pt+.5\pgflinewidth}
    \begin{pgfscope}
      \pgfpathmoveto{\pgfpoint{0pt}{\lw}}%
      \pgfpathlineto{\pgfpoint{\pgfdecoratedinputsegmentlength}{\lw}}%
      \pgfmathtruncatemacro\dashnum{%
        round((\pgfdecoratedinputsegmentlength-3pt)/6pt)
      }
      \pgfmathsetmacro\dashscale{%
        \pgfdecoratedinputsegmentlength/(\dashnum*6pt + 3pt)
      }
      \pgfmathsetlengthmacro\dashunit{3pt*\dashscale}
      \pgfsetdash{{\dashunit}{\dashunit}}{0pt}
      \pgfusepath{stroke}
      \pgfsetdash{}{0pt}
      \pgfpathmoveto{\pgfpoint{0pt}{-\lw}}%
      \pgfpathlineto{\pgfpoint{\pgfdecoratedinputsegmentlength}{-\lw}}%
      \pgfusepath{stroke}
    \end{pgfscope}
  }
}

\begin{document}

\hypersetup{pageanchor=false}
\begin{titlepage}

\begin{center}

\hfill CERN-TH-2025-117 \\
\hfill RESCEU-14/25 \\
\hfill IPMU25-0030\\
\hfill KOBE-COSMO-25-13 \\
\hfill KEK-TH-2731 \\

\vskip 0.5in

{\Huge \bfseries Cancellation of one-loop correction to \vspace{5mm} \\ soft tensor power spectrum
} \\
\vskip .8in

{\Large Yohei Ema$^{a}$, Muzi Hong$^{b}$, Ryusuke Jinno$^{c}$, Kyohei Mukaida$^{d}$}

\vskip .3in
\begin{tabular}{ll}
$^{a}$ &\!\!\!\!\!\emph{Theoretical Physics Department, CERN, 1211 Geneva 23, Switzerland}\\
$^{b}$ &\!\!\!\!\!\emph{Department of Physics, Graduate School of Science, The University of Tokyo, }\\[-.15em]
& \!\!\!\!\!\emph{Tokyo 113-0033, Japan}\\
$^{b}$ &\!\!\!\!\!\emph{RESCEU, Graduate School of Science, The University of Tokyo, Tokyo 113-0033, Japan} \\
$^{b}$ &\!\!\!\!\!\emph{Kavli IPMU (WPI), UTIAS, The University of Tokyo, Kashiwa 277-8583, Japan}\\
$^{c}$ &\!\!\!\!\!\emph{Department of Physics, Graduate School of Science, Kobe University,}\\[-.15em]
& \!\!\!\!\!\emph{1-1 Rokkodai, Kobe, Hyogo 657-8501, Japan} \\
$^{d}$ & \!\!\!\!\!\emph{Theory Center, IPNS, KEK, 1-1 Oho, Tsukuba, Ibaraki 305-0801, Japan}\\
$^{d}$ & \!\!\!\!\!\emph{Graduate University for Advanced Studies (Sokendai),}\\[-.15em]
& \!\!\!\!\!\emph{1-1 Oho, Tsukuba, Ibaraki 305-0801, Japan}
\end{tabular}

\end{center}
\vskip .6in

\begin{abstract}
\noindent
We demonstrate that there are no scale-invariant one-loop corrections to the superhorizon tensor perturbations from small-scale (potentially enhanced) scalar perturbations, irrespective of the details of inflationary background time evolution.
For this purpose we derive a soft tensor effective field theory at leading order in the gradient expansion by integrating out small-scale scalar fluctuations in a general time-dependent background over the Schwinger--Keldysh contour, \textit{i.e.,} we perform loop calculations in the soft limit of external momentum.
The absence of scale-invariant corrections originates from the diffeomorphism invariance of general relativity and is therefore unavoidable.
\end{abstract}

\end{titlepage}

\tableofcontents
\renewcommand{\thepage}{\arabic{page}}
\renewcommand{\thefootnote}{$\natural$\arabic{footnote}}
\setcounter{footnote}{0}
\hypersetup{pageanchor=true}

\section{Introduction}
\label{sec:introduction}

Inflationary cosmology predicts the existence of primordial perturbations in the early Universe.
One of the characteristic features of these perturbations is a nearly scale-invariant spectrum in perfect agreement with the large-scale observations, such as cosmic microwave background (CMB) anisotropies~\cite{Planck:2018vyg,ACT:2025fju}
and the large-scale structure (LSS)~\cite{BOSS:2016wmc,DESI:2024mwx,DESI:2024hhd}.
In addition to the scalar perturbations, inflationary cosmology predicts tensor perturbations, corresponding to the primordial gravitational waves, whose null observations put upper bounds on the inflationary scale~\cite{Planck:2018jri,BICEP:2021xfz,ACT:2025tim}.

In contrast, cosmological perturbations on small scales are not well constrained, leaving room to potentially generate interesting imprints in the current Universe.
If large curvature perturbations exist on small scales, these enhanced perturbations may collapse into primordial black holes~\cite{Zeldovich:1967lct} (PBHs) after the horizon re-entry.
After the discovery of gravitational waves from the BH merger in the LIGO/Virgo detector~\cite{LIGOScientific:2016aoc}, PBHs have been drawing renewed attention as a candidate for the LIGO/Virgo events as well as that for dark matter~\cite{Chapline:1975ojl} 
(see, \textit{e.g.}, Refs.~\cite{Carr:2020gox,Carr:2020xqk,Green:2020jor} for reviews).
Moreover, such enhanced scalar perturbations generate tensor perturbations by a second-order effect known as induced gravitational waves, which can be probed by future gravitational wave detectors~\cite{LISA:2017pwj,Punturo:2010zz,LIGOScientific:2016wof,Ruan:2018tsw,Wang:2019ryf,Kawamura:2020pcg}.

Recently, partly motivated by these activities, there have been growing interests in the one-loop corrections to the primordial perturbations.
A couple of recent studies have claimed that such enhanced small-scale perturbations have a significant impact on the large-scales ones in the context of transient ultra-slow-roll inflation models~\cite{Kristiano:2022maq,Kristiano:2023scm,Ota:2022hvh,Ota:2022xni}.
This is quite surprising, if true, since the production of small-scale perturbations occurs at the time of their horizon crossing that occurs well after that of the large-scale modes.
Hence one would expect the $k^3$ suppression for the power spectrum from causality, implying the decoupling of the small-scale perturbations, \emph{i.e.}, the small-scale perturbations do not affect the large-scale ones in the soft limit $k \to 0$.
The induced gravitational waves at the second order mentioned above obey this causality suppression.
Indeed, it has been already shown (see, \emph{e.g.}, Refs.~\cite{Senatore:2009cf,Pimentel:2012tw,Senatore:2012ya,Assassi:2012et,Tanaka:2015aza})
that the loop corrections of small-scale perturbations to the large-scale scalar perturbations are absent in single-clock inflation models, prior to the recent studies on ultra-slow-roll inflation and PBHs.
Nevertheless, Refs.~\cite{Kristiano:2022maq,Kristiano:2023scm} and Refs.~\cite{Ota:2022hvh,Ota:2022xni} have challenged this understanding, claiming the existence of scale-invariant corrections from the small-scale scalar perturbations to the large-scale scalar and tensor perturbations,
respectively, that survive in the soft limit.
This claim has ignited huge follow-up investigations of the one-loop corrections to the superhorizon scalar perturbations~\cite{
Riotto:2023hoz,Choudhury:2023vuj,Choudhury:2023jlt,Riotto:2023gpm,Choudhury:2023rks,Firouzjahi:2023aum,Motohashi:2023syh,Firouzjahi:2023ahg,Franciolini:2023agm,Cheng:2023ikq,Fumagalli:2023hpa,Maity:2023qzw,Tada:2023rgp,Firouzjahi:2023bkt,Iacconi:2023ggt,Inomata:2024lud,Kawaguchi:2024rsv,Ballesteros:2024zdp,Kristiano:2024ngc,Fumagalli:2024jzz,Sheikhahmadi:2024peu,Inomata:2025bqw,Fang:2025vhi,Inomata:2025pqa,Braglia:2025cee,Braglia:2025qrb,Kristiano:2025ajj},
while only a limited number of studies exist in the case of the tensor perturbations at this moment.\footnote{
	Ref.~\cite{Firouzjahi:2023btw} claims that the correction from the USR phase 
	to the superhorizon tensor mode is small (but non-zero), while we will see that the correction actually vanishes in the soft limit.
}

In this paper, contrary to the aforementioned claim, we show with an explicit calculation the absence of scale invariant one-loop corrections from the small-scale scalar perturbations to the large-scale \emph{tensor} perturbations in the soft limit.
To this end, we consider a spectator scalar field minimally coupled to gravity with time-dependent sound speed and mass. 
We also take an overall coefficient of the spectator field Lagrangian as an arbitrary function of time instead of the scale factor in de~Sitter spacetime so that it can cover more scenarios (see Eq.~\eqref{eq-model} and below).
Our calculation is independent of the details of the inflationary background evolution, and therefore applies to, among others, the transient ultra-slow-roll models frequently discussed in the literature in the context of the enhanced curvature perturbation at small scales.
To understand the soft limit, all we need is the effective field theory (EFT) of super-horizon tensor modes, dubbed as \textit{soft} tensor EFT.
By explicitly integrating out the scalar perturbations, namely the \textit{hard} scalar modes, on the Schwinger--Keldysh contour, we demonstrate that the soft tensor EFT never contains the term responsible for the scale-invariant enhancement.
In other words, we calculate the one-loop corrections from the spectator field in the soft limit of external momenta, and show the cancellation (at the level of the conformal time integrand; see the discussion at the end of Sec.~\ref{sec:ra_basis}).
Our explicit calculation highlights the importance of an identity involving Green's functions (see Eq.~\eqref{eq:Grr_logl_deriv} below) for the cancellation, which holds for an arbitrary inflationary background evolution as long as one solves the mode equations in a given time-dependent background consistently.
We remark that this conclusion stems from the diffeomorphism invariance of general relativity.

The rest of the paper is organized as follows.
In Sec.~\ref{sec:ra_basis}, we briefly review the in-in formalism to define the notation, with a particular emphasis on the Keldysh $r/a$-basis used throughout this paper.
Sec.~\ref{sec:one_loop} is the main part.
We show explicitly the cancellation of the one-loop corrections to the tensor perturbations in the soft limit by integrating out a spectator scalar field in Sec.~\ref{subsec:spectator_field}. 
We then remark its connection to the diffeomorphism invariance of general relativity in Sec.~\ref{subsec:symmetry}.
Finally, we conclude and comment on several possible future directions in Sec.~\ref{sec:conclusion}.

\section{Mini-review: in-in formalism and Keldysh $r/a$-basis}
\label{sec:ra_basis}

\begin{figure}[t]
	\centering
 	\includegraphics[width=0.45\linewidth]{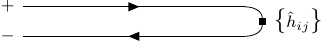}
	\vspace{2.5mm}
	\caption{The path integral contour of the in-in formalism for the cosmological correlators.
	We evaluate the correlation functions of gravitons at the future infinity, which corresponds to the end of inflation.}
	\label{fig:in-in_contour}
\end{figure}

In the context of primordial cosmology,
we evaluate the cosmological correlators at the end of inflation, which we take at the conformal time $\tau = 0$ with the initial condition fixed.\footnote{
	We take the Bunch-Davies vacuum throughout this paper.
}
Correlation functions are then calculated by the in-in formalism~\cite{Schwinger:1960qe,Bakshi:1962dv,Bakshi:1963bn,Keldysh:1964ud}.
In the following, we briefly summarize the basic properties of the in-in formalism necessary for our calculation.
In particular, we introduce the so-called Keldysh $r/a$-basis~\cite{Keldysh:1964ud} used throughout the paper.
For a more comprehensive review on the topic, see \textit{e.g.}, Refs.~\cite{Caron-Huot:2007zhp,Laine:2016hma,Ghiglieri:2020dpq} and references therein.

The in-in formalism contains two distinct path integral contours shown in Fig.~\ref{fig:in-in_contour}, and correspondingly, we have four Green's functions:\footnote{
	Here and in what follows, we focus on bosonic degrees of freedom.
	It can be trivially extended to fermions after assigning the fermionic minus signs appropriately.
}
\begin{align}
	\langle \mathcal{T}_\mathcal{C}\, \phi_+(x) \phi_+(y)\rangle 
	&= G_{F}(x,y) = \theta(x^0-y^0)\langle \phi(x) \phi(y)\rangle + \theta(y^0-x^0)\langle \phi(y) \phi(x)\rangle,
	\\
	\langle \mathcal{T}_\mathcal{C}\, \phi_-(x) \phi_-(y)\rangle 
	&= G_{\bar{F}}(x,y) = \theta(y^0-x^0)\langle \phi(x) \phi(y)\rangle + \theta(x^0-y^0)\langle \phi(y) \phi(x)\rangle,
	\\
	\langle \mathcal{T}_\mathcal{C}\, \phi_+(x) \phi_-(y)\rangle 
	&= G_{<}(x,y) = \langle \phi(y) \phi(x) \rangle,
	\quad
	\langle \mathcal{T}_\mathcal{C}\, \phi_-(x) \phi_+(y)\rangle 
	= G_{>}(x,y) = \langle \phi(x) \phi(y) \rangle,
\end{align}
where the fields $\phi_\pm$ originate from the $\pm$ contours respectively, and $\mathcal{T}_\mathcal{C}$ denotes the path ordering along the complex-time contour given in Fig.~\ref{fig:in-in_contour}.
Interaction vertices arise either from the $+$ or $-$ contour,
and they have the opposite overall sign due to the opposite direction of the path integral contours.
Each contribution to the correlators is then expressed by the in-in Feynman diagrams.

Although one can compute the correlators in the original $\pm$ basis, 
it is more convenient to move to another field basis, called the Keldysh $r/a$-basis,
by taking a linear transformation of the fields as
\begin{align}
	\phi_r = \frac{1}{2}(\phi_+ + \phi_-),
	\quad
	\phi_a = \phi_+ - \phi_-.
\end{align}
In this basis, the two-point correlators are given by
\begin{align}
	\langle \mathcal{T}_\mathcal{C}\, \phi_r(x) \phi_r(y)\rangle
	&= G_{rr}(x,y) = \frac{1}{2}\left[G_{>}(x,y) + G_{<}(x,y)\right],
	\quad
	\langle \mathcal{T}_\mathcal{C}\, \phi_a(x) \phi_a(y)\rangle = 0,
	\\
	\langle \mathcal{T}_\mathcal{C}\, \phi_r(x) \phi_a(y)\rangle
	&= \langle \mathcal{T}_\mathcal{C}\, \phi_a(y) \phi_r(x)\rangle = G_{ra}(x,y) = \theta(x^0-y^0)\left[G_{>}(x,y) - G_{<}(x,y)\right].
\end{align}
Thus, the $ra$-correlator is given by the retarded function,
while the $rr$-correlator does not contain the time-ordering theta function and is called the statistical function.
The $aa$-correlator vanishes identically and this is the simplest example of a general theorem that arbitrary correlators
of only $a$-type operators vanish identically.
To make the causal structure manifest, it is convenient to assign the arrow, representing the causal time flow, as
\begin{align}
    G_{rr} (x,y) &=
    \begin{tikzpicture}[baseline=(c)]
        \begin{feynman}[inline = (base.c), every blob = {/tikz/fill=gray!30,/tikz/inner sep = 2pt}]
            \vertex [label=\({\scriptstyle \phi_r (x)}\)](f1);
            \vertex [right = 1.2 of f1,label=\({\scriptstyle \phi_r(y)}\)] (v1);
            \vertex [below = 0.1 of f1] (c);
            \draw (f1) -- node[midway]{$\parallel$} (v1);
            \diagram*{
            (f1) -- [anti majorana] (v1),
            };
        \end{feynman}
    \end{tikzpicture}\,, \qquad
    G_{ra} (x,y) =
    \begin{tikzpicture}[baseline=(c)]
        \begin{feynman}[inline = (base.c), every blob = {/tikz/fill=gray!30,/tikz/inner sep = 2pt}]
            \vertex [label=\({\scriptstyle \phi_r(x)}\)](f1);
            \vertex [right = 1.2 of f1,label=\({\scriptstyle \phi_a(y)}\)] (v1);
            \vertex [below = 0.1 of f1] (c);
            \diagram*{
            (f1) -- [anti fermion] (v1),
            };
        \end{feynman}
    \end{tikzpicture}\,.
\end{align}
In this notation, the causal arrow flows out of $a$-fields and flows into $r$-fields.
Since the in-in action takes the form of $S_+ - S_-$ in a unitary theory, and hence is antisymmetric with respect to the exchange $+ \leftrightarrow -$, each vertex contains an odd number of $a$-fields.
Thus an odd number of arrows come out of each vertex in the Feynman diagrams.
It is also clear that a diagram with a closed loop of the time arrow identically vanishes due to causality.
The fields on the end-of-inflation surface lie at the boundary of the $\pm$ contours, and they are interpreted as $r$-fields.
This means that, in each Feynman diagram, the causal time arrows eventually flow into the boundary fields.

If we consider the tensor power spectrum, in general, we have the following contributions
\begin{align}
	\begin{tikzpicture}[baseline = (c)]
		\begin{feynman}[inline = (base.c)]
			\vertex (f1);
			\vertex [right = 0.5cm of f1] (c1);
			\vertex [right = 0.75cm of c1] (m1);
			\vertex [right = 1.5cm of c1] (c2);
			\vertex [right = 0.5cm of c2] (f2);
			\vertex [below = of m1, blob, shape = ellipse, minimum height = 0.75cm, minimum width = 1.5cm] (v1){};
			\vertex [left = 0.75cm of v1, square dot](v1p){};
			\vertex [right = 0.75cm of v1, square dot](v2p){};
			\vertex [below = 0.7 of c1] (c);
			\diagram*{
			(f1) -- [very thick, darkred] (f2),
			(c1) -- [photon](v1p),
			(c2) -- [photon](v2p),
			};
		\end{feynman}
	\end{tikzpicture}
	\,=\,
	\begin{tikzpicture}[baseline = (c)]
		\begin{feynman}[inline = (base.c), every blob = {/tikz/fill=gray!30,/tikz/inner sep = 2pt}]
			\vertex (f1);
			\vertex [right = 0.5cm of f1] (c1);
			\vertex [right = 0.75cm of c1] (m1);
			\vertex [left = 0.2cm of m1] (m3);
			\vertex [right = 1.5cm of c1] (c2);
			\vertex [right = 0.5cm of c2] (f2);
			\vertex [below = of m1, blob, shape = ellipse, minimum height = 0.75cm, minimum width = 1.5cm] (v1){$\longrightarrow$};
			\vertex [left = 0.75cm of v1, square dot](v1p){};
			\vertex [right = 0.75cm of v1, square dot](v2p){};
			\vertex [below = 0.7cm of c1] (c);
			\vertex [below = 0.65cm of f1] (m0);
			\vertex [below = 0.65cm of m3] (m2);
			\vertex [below = 0.65cm of c1] (c1c);
			\vertex [below = 0.649cm of c2] (c2c1);
			\vertex [below = 0.651cm of c2] (c2c2);
			\vertex [below = 0.899cm of c1] (c1c1);
			\vertex [below = 0.901cm of c1] (c1c2);
			\vertex [below = 0.399cm of c1] (c1c3);
			\vertex [below = 0.401cm of c1] (c1c4);
			\draw (c1c) -- node[midway, rotate = 90]{$\parallel$} (c1c);
			\diagram*{
			(f1) -- [very thick, darkred] (f2),
			(c1) -- [photon] (v1p),
			(c2) -- [photon] (v2p),
			(c2c2) -- [fermion] (c2c1),
			(c1c1) -- [fermion] (c1c2),
			(c1c4) -- [fermion] (c1c3),
			};
		\end{feynman}
	\end{tikzpicture}
	\,+\,
	\begin{tikzpicture}[baseline = (c)]
		\begin{feynman}[inline = (base.c), every blob = {/tikz/fill=gray!30,/tikz/inner sep = 2pt}]
			\vertex (f1);
			\vertex [right = 0.5cm of f1] (c1);
			\vertex [right = 0.75cm of c1] (m1);
			\vertex [right = 0.2cm of m1] (m3);
			\vertex [right = 1.5cm of c1] (c2);
			\vertex [right = 0.5cm of c2] (f2);
			\vertex [below = of m1, blob, shape = ellipse, minimum height = 0.75cm, minimum width = 1.5cm] (v1){$\longleftarrow$};
			\vertex [left = 0.75cm of v1, square dot](v1p){};
			\vertex [right = 0.75cm of v1, square dot](v2p){};
			\vertex [below = 0.7cm of c1] (c);
			\vertex [below = 0.65cm of f2] (m0);
			\vertex [below = 0.65cm of m3] (m2);
			\vertex [below = 0.65cm of c2] (c2c);
			\vertex [below = 0.899cm of c2] (c2c1);
			\vertex [below = 0.901cm of c2] (c2c2);
			\vertex [below = 0.399cm of c2] (c2c3);
			\vertex [below = 0.401cm of c2] (c2c4);
			\vertex [below = 0.649cm of c1] (c1c1);
			\vertex [below = 0.651cm of c1] (c1c2);
			\draw (c2c) -- node[midway, rotate = 90]{$\parallel$} (c2c);
			\diagram*{
			(f1) -- [very thick, darkred] (f2),
			(c1) -- [photon] (v1p),
			(c1c2) -- [fermion] (c1c1),
			(c2c1) -- [fermion] (c2c2),
			(c2c4) -- [fermion] (c2c3),
			(c2) -- [photon] (v2p),
			};
		\end{feynman}
	\end{tikzpicture}
	\, + \,
	\begin{tikzpicture}[baseline=(c)]
		\begin{feynman}[inline = (base.c), every blob = {/tikz/fill=gray!30,/tikz/inner sep = 2pt}]
			\vertex (f1);
			\vertex [right = 0.5cm of f1] (c1);
			\vertex [right = 0.75cm of c1] (m1);
			\vertex [right = 0.2cm of m1] (m3);
			\vertex [right = 0.75cm of c1] (c1c2);
			\vertex [below = 0.79cm of c1c2] (v1v2_1);
			\vertex [below = 0.99cm of c1c2] (v1v2_2);
			\vertex [below = 1.69cm of c1c2] (v1v2_3);
			\vertex [right = 1.5cm of c1] (c2);
			\vertex [right = 0.5cm of c2] (f2);
			\vertex [right = 0.25cm of c1](mv1);
			\vertex [left = 0.25cm of c2](mv2);
			\vertex [below = of mv1, blob, minimum height = 0.5cm, minimum width = 0.5cm] (v1){$\shortleftarrow$};
			\vertex [below = of mv2, blob, minimum height = 0.5cm, minimum width = 0.5cm] (v2){$\shortrightarrow$};
			\vertex [above = 0cm of c1c2](v1v2);
			\vertex [left = 0.25cm of v1, square dot](v1p){};
			\vertex [right = 0.25cm of v2, square dot](v2p){};
			\vertex [below = 0.7cm of c1] (c);
			\vertex [below = 0.6cm of m1] (m0);
			\vertex [below = 1.35cm of m0] (m2);
			\vertex [left = 0.12cm of m2] (vdots);
			\node [above = 0.32cm of vdots, steelblue] {\vdots};
			\vertex [below = 0.599cm of c1] (c1c1);
			\vertex [below = 0.601cm of c1] (c1c2);
			\vertex [below = 0.599cm of c2] (c2c1);
			\vertex [below = 0.601cm of c2] (c2c2);
			\diagram*{
			     (f1) -- [very thick, darkred] (f2),
			     (c1) -- [photon] (v1p),
			     (c1c2) -- [fermion] (c1c1),
			     (c2c2) -- [fermion] (c2c1),
			     (c2) -- [photon] (v2p),
			     (v1v2_1) -- [out = 180, in = 70, fermion, steelblue] (v1),
			     (v1v2_1) -- [out = 0, in = 110, fermion, steelblue] (v2),
			     (v1v2_2) -- [out = 180, in = 40, fermion, steelblue] (v1),
			     (v1v2_2) -- [out = 0, in = 140, fermion, steelblue] (v2),
			     (v1v2_3) -- [out = 180, in = 300, fermion, steelblue] (v1),
			     (v1v2_3) -- [out = 0, in = 240, fermion, steelblue] (v2),
			(m0) -- [dashed, steelblue, very thick] (m2),
			};
		\end{feynman}
	\end{tikzpicture}
        \,,
    \label{eq:tensor_2pt_general_diagram}
\end{align}
where the red thick line corresponds to the end-of-inflation hypersurface, the wavy lines are the bulk-to-boundary graviton propagators,
and the blob indicates the bulk dynamics, with the arrow corresponding to the direction of the causal flow.\footnote{
	If we use the statistical propagators for both sides of the bulk-to-boundary propagators,
	it vanishes since the bulk part corresponds to the correlation functions of only $a$-type operators.
	In the one-loop case we consider below, one can easily confirm this since the diagram necessarily forms a closed causal time loop.
}
In the last diagram, we have used the cutting rule for in-in correlators~\cite{Ema:2024hkj} where the dashed line corresponds to the location of the cut.
Here we draw the diagrams as if the two external gravitons are connected to different vertices, but they can be connected to the same vertex, in which case we do not have the last diagram.

In the following, we focus on the former two contributions since only these diagrams potentially contribute to the power spectrum in the soft limit~\cite{Tada:2023rgp}.
Since the graviton satisfies the same mode equation as the massless scalar field in the inflationary background, the bulk-to-boundary propagators of the graviton are given by essentially the same expressions as those of the massless scalar field in de~Sitter spacetime, which are given, after going to the momentum space, by\footnote{
The normalization is different, which is irrelevant to later discussion.
}
\begin{align}
	\Delta_{rr}(k;0,\tau) &= \frac{H^2}{2k^3}\left[\cos(k\tau) + k\tau \sin(k\tau)\right],
	\quad
	\Delta_{ra}(k;0,\tau) = \frac{iH^2}{k^3}\left[\sin(k\tau)-k\tau\cos(k\tau)\right],
\end{align}
where $H$ is the Hubble parameter, the conformal time $\tau$ spans $[-\infty,0]$ during inflation,
and $k = \vert \vec{k}\vert$ is the size of the three-momentum $\vec{k}$.
In the soft limit, they reduce to
\begin{align}
	\lim_{k\to0}\Delta_{rr}(k;0,\tau) = \frac{H^2}{2k^3},
	\quad
	\lim_{k\to0}\Delta_{ra}(k;0,\tau) = \frac{iH^2\tau^3}{3}.
    \label{eq:soft_limit_bulk-to-boundary}
\end{align}
Therefore, combined with the factor $k^3/2\pi^2$ arising from the momentum integral measure in the definition of the power spectrum,
only those diagrams with the bulk-to-boundary statistical propagator, $\Delta_{rr}$, potentially survive in the soft limit of the boundary fields.
On the other hand, the last diagram in Eq.~\eqref{eq:tensor_2pt_general_diagram}
contains the standard contribution to the gravitational wave induced at the second order
by the scalar perturbations, which is suppressed by $k^3$ in the superhorizon limit.

Before moving on, let us note one caveat. In the above argument, and also in the following, we focus on $\mathcal{O}(k^{-3})$ terms in the soft limit 
\emph{before} performing the conformal time integral of the interaction vertices, \emph{i.e.}, at the level of the integrand. 
In addition, sub-leading order terms in the soft limit at the integrand level
can nevertheless behave as $\mathcal{O}(k^{-3})$ \emph{after} the conformal time integral, by picking up additional powers of $k^{-1}$ 
from the conformal time integral.
The wavefunction renormalization could be one such example.
These terms are expected to arise from the integral in the infinite past region for a non-singular integrand,\footnote{
	This can be understood as follows. For a finite $k$, the $e^{\pm ik\tau}$ factor in the integrand, together with the $i\epsilon$ prescription,
	results in a finite value for the integral region $\tau \to -\infty$ due to its oscillatory feature.
	However, this oscillation goes away in the limit $k\to 0$, resulting in a non-cancellation and hence a singularity.
	Indeed, Paley-Wiener-Schwartz theorem guarantees that a Fourier transform of a non-singular function with a compact support is an entire function,
	meaning that the singular behavior should arise from the boundaries of the conformal time integrals.
}
and are expected to be insensitive to the dynamics at the intermediate time scale such as the transient USR phase.
Therefore, in the following, we focus only on the $\mathcal{O}(k^{-3})$ terms at the integrand level,
which are potentially sensitive to the dynamics at the intermediate time scale.
It would be instructive to check this argument, 
that the $\mathcal{O}(k^{-3})$ term after the conformal time integral is sensitive only to the infinite past,
with an explicit example. We leave it as a future work.\footnote{
	For instance, Ref.~\cite{Ballesteros:2024cef} obtained a logarithmic correction to the tensor power spectrum in de~Sitter spacetime,
	which presumably arises from the sub-leading order terms at the integrand level.
	It would be interesting to see if the same correction is obtained for a model with the transient USR phase,
	as long as the cosmology in the infinite past is the same.
}

\section{One-loop correction to soft primordial gravitons}
\label{sec:one_loop}

We now calculate the one-loop correction of a scalar field to the tensor power spectrum in the soft limit of the external gravitons.
This is expected to vanish since, if exists, it corresponds to an operator in the in-in effective field theory (obtained after integrating out the scalar fields)
of the form
\begin{align}
	\mathcal{L} \sim h_{a ij}h_r^{ij} + \cdots,
\end{align}
which is forbidden by the general coordinate transformation invariance (see Sec.~\ref{subsec:symmetry}).\footnote{
	On the other hand, the scale invariant corrections \emph{after} the conformal time integral correspond to, \emph{e.g.},
	the graviton kinetic term that is allowed by the symmetry.
}
In the following, we confirm this expectation explicitly;
we show that the one-loop correction of a scalar field to the primordial tensor power spectrum vanishes in the soft limit for an arbitrary time evolution of the background.\footnote{
	Cancellations of cosmological loop corrections in general have been already extensively studied in Refs.~\cite{Senatore:2009cf,Pimentel:2012tw,Senatore:2012ya,Assassi:2012et,Tanaka:2015aza,Fumagalli:2023hpa,Tada:2023rgp,Inomata:2024lud,Kawaguchi:2024rsv,Fumagalli:2024jzz,Inomata:2025bqw,Fang:2025vhi,Inomata:2025pqa,Braglia:2025cee,Braglia:2025qrb}, and also their relation to the Ward--Takahashi identity~\cite{Assassi:2012et,Tanaka:2015aza}, mostly based on the operator formalism.
	Our analysis can be regarded as a complementary approach based on the path integral formalism on the Schwinger--Keldysh contour, and focuses on the cancellation of the small-scale loop corrections to the large-scale tensor perturbations in the soft limit.
	Also, our analysis is independent of a specific mechanism to enhance the small-scale perturbations, and therefore applies to a wide class of models.
}

\subsection{Spectator field}
\label{subsec:spectator_field}

We consider the one-loop correction from a spectator scalar field $\chi$. We take the action as
\begin{align}
	S = \frac{1}{2}\int \dd^{d-1}x \dd\tau\,f^2(\tau) \left[\left(\frac{\dd\chi}{\dd\tau}\right)^2 - c_s^2 (\tau)\left[e^{-h}\right]_{ij}(\partial_i \chi)(\partial_j \chi)
	- m_\chi^2 (\tau) \chi^2
	\right].
    \label{eq-model}
\end{align}
Here $c_s(\tau)$ is the sound speed and $m_\chi (\tau)$ is the mass term, both of which can depend on time.
The graviton $h_{ij}$ satisfies the transverse-traceless condition,
and $d$ is the spacetime dimension which we take to be arbitrary at this moment.
In the narrow sense of spectator fields, the time-dependent function $f(\tau)$ is usually taken to be equal to the scale factor.
However, we do not impose this restriction here so as to allow for a (transient) phase that enhances $\chi$ perturbations on small scales such as the axion-curvaton model~\cite{Kawasaki:2012wr} or $f(\tau) \propto \tau^2$ during the ultra-slow-roll phase~\cite{Kinney:1997ne,Inoue:2001zt,Kinney:2005vj,Martin:2012pe,Motohashi:2017kbs}.\footnote{
In this case, $\chi$ models adiabatic perturbation.
}
We also take the time-dependence of $c_s (\tau)$, and $m_\chi (\tau)$ to be arbitrary, which allows for enhanced $\chi$ perturbations on small scales by other mechanisms~\cite{Yokoyama:1995ex,Chen:2019zza}.

Our purpose here is to calculate the one-loop correction of the spectator field $\chi$ to the tensor power spectrum in the soft limit.
In particular, we show the cancellation of the one-loop corrections in the soft limit, keeping $f(\tau)$, $c_s (\tau)$, and $m_\chi (\tau)$ arbitrary functions of $\tau$.
In a realistic situation, $f(\tau)$, $c_s (\tau)$, and $m_\chi (\tau)$ are related to the background inflationary dynamics, and therefore our proof of the cancellation does not depend on the details of the background dynamics, and applies to, among others, the case with a transient ultra-slow-roll phase between the standard slow-roll inflationary phases.

As we have explained below Eq.~\eqref{eq:tensor_2pt_general_diagram}, we focus on the diagrams with the bulk-to-boundary statistical propagators.
In the $r/a$-basis, the relevant in-in interaction terms are given by
\begin{align}
	S_{\mathrm{in}\mathchar`-\mathchar`-\mathrm{in}} = 
	\int \dd^{d-1}x \dd\tau\,f^2(\tau) c_s^2(\tau) \left[\frac{1}{2}\hat{h}_{a}^{ij} (\partial_i \hat{\chi}_r) (\partial_j\hat{\chi}_r)
	+ \hat{h}_{r}^{ij} (\partial_i \hat{\chi}_r)(\partial_j \hat{\chi}_a)
	-\frac{1}{2}\hat{h}_{ak}^{i} \hat{h}_{r}^{jk} (\partial_i \hat{\chi}_r) (\partial_j\hat{\chi}_r)
	\right]
	+ \cdots,
\end{align}
where we put the hats to indicate that these fields are operators, 
and the spatial indices are now contracted by $\delta_{ij}$.
At one loop, the relevant diagrams are given by
\begin{align}
	\left[
	\begin{tikzpicture}[baseline=(middle)]
		\begin{feynman}[inline = (base.middle), every blob = {/tikz/fill=gray!30,/tikz/inner sep = 2pt}]
			\vertex (f1);
			\vertex [right = 0.5cm of f1] (c1);
			\vertex [right = 0.75cm of c1] (m1);
			\vertex [right = 0.2cm of m1] (m3);
			\vertex [right = 1.5cm of c1] (c2);
			\vertex [right = 0.5cm of c2] (f2);
			\vertex [below = of m1, blob, shape = ellipse , minimum height = 0.75cm, minimum width = 1.5cm] (v1){$\longleftarrow$};
			\vertex [left = 0.75cm of v1, square dot](v1p){};
			\vertex [right = 0.75cm of v1, square dot](v2p){};
			\vertex [below = 0.7cm of c1] (c);
			\vertex [below = 0.65cm of f2] (m0);
			\vertex [below = 0.65cm of m3] (m2);
			\vertex [below = 0.65cm of c2] (c2c);
			\vertex [below = 0.899cm of c2] (c2c1);
			\vertex [below = 0.901cm of c2] (c2c2);
			\vertex [below = 0.399cm of c2] (c2c3);
			\vertex [below = 0.401cm of c2] (c2c4);
			\vertex [below = 0.649cm of c1] (c1c1);
			\vertex [below = 0.651cm of c1] (c1c2);
			\draw (c2c) -- node[midway, rotate = 90]{$\parallel$} (c2c);
			\vertex [below = 0.9 of f1] (middle);
			\diagram*{
			(f1) -- [very thick, darkred] (f2),
			(c1) -- [photon] (v1p),
			(c1c2) -- [fermion] (c1c1),
			(c2c1) -- [fermion] (c2c2),
			(c2c4) -- [fermion] (c2c3),
			(c2) -- [photon] (v2p),
			};
		\end{feynman}
	\end{tikzpicture}
	\right]_{1\mathchar`-\mathchar`-\mathrm{loop}}
	&=
	\begin{tikzpicture}[baseline=(middle)]
		\begin{feynman}[inline = (base.middle), every blob = {/tikz/fill=gray!30,/tikz/inner sep = 2pt}]
			\vertex (f1);
			\vertex [below = 0.9cm of f1] (middle);
			\vertex [right = 0.5cm of f1] (c1);
			\vertex [right = 0.75cm of c1] (m1);
			\vertex [right = 1.5cm of c1] (c2);
			\vertex [right = 0.5cm of c2] (f2);
			\vertex [below = 1.25cm of m1] (v1);
			\vertex [left = 0.45cm of v1](v1p);
			\vertex [right = 0.45cm of v1](v2p);
			\vertex [below = 0.7cm of c1] (c);
			\vertex [below = 0.4cm of v1] (vu);
			\draw (vu) -- node[midway]{$\parallel$} (vu);
			\vertex [below left = 0.625cm and 0.15cm of c2] (m2);
			\draw (m2) -- node[midway, rotate = 76.5043]{$\parallel$} (m2); 
			\vertex [above right = 0.25cm and 0.06cm of m2] (m2u);
			\vertex [above right = 0.0125cm and 0.003cm of m2u] (m2u1);
			\vertex [below left = 0.0125cm and 0.003cm of m2u] (m2u2);
			\vertex [below left = 0.25cm and 0.06cm of m2] (m2d);
			\vertex [above right = 0.0125cm and 0.003cm of m2d] (m2d1);
			\vertex [below left = 0.0125cm and 0.003cm of m2d] (m2d2);
			\vertex [below right = 0.625cm and 0.15cm of c1] (m3);
			\vertex [above left = 0.0125cm and 0.003cm of m3] (m31);
			\vertex [below right = 0.0125cm and 0.003cm of m3] (m32);
			\diagram*{
			(f1) -- [very thick, darkred] (f2),
			(v2p) -- [half right, fermion] (v1p) -- [half right, anti majorana] (v2p),
			(c1) -- [photon] (v1p),
			(c2) -- [photon] (v2p),
			(m2u2) -- [fermion] (m2u1),
			(m2d1) -- [fermion] (m2d2),
			(m32) -- [fermion] (m31),
			};
		\end{feynman}
	\end{tikzpicture}
	\,+\,
	\begin{tikzpicture}[baseline=(middle)]
		\begin{feynman}[inline = (base.middle), every blob = {/tikz/fill=gray!30,/tikz/inner sep = 2pt}]
			\vertex (f1);
			\vertex [below = 0.9cm of f1] (middle);
			\vertex [right = 0.5cm of f1] (c1);
			\vertex [right = 0.75cm of c1] (m1);
			\vertex [right = 1.5cm of c1] (c2);
			\vertex [right = 0.5cm of c2] (f2);
			\vertex [below = of m1] (v1);
			\vertex [below = 0.75cm of v1] (v2);
			\vertex [below = 0.7cm of c1] (c);
			\vertex [below left = 0.5cm and 0.375cm of c2] (c2c);
			\vertex [below right = 0.5cm and 0.375cm of c1] (c1c);
			\vertex [above left = 0.002cm and 0.0015cm of c1c] (c1c1);
			\vertex [below right = 0.002cm and 0.0015cm of c1c] (c1c2);
			\vertex [above right = 0.2cm and 0.15cm of c2c] (m1);
			\vertex [above right = 0.002cm and 0.0015cm of m1] (m11);
			\vertex [below left = 0.002cm and 0.0015cm of m1] (m12);
			\vertex [below left = 0.2cm and 0.15cm of c2c] (m2);
			\vertex [above right = 0.002cm and 0.0015cm of m2] (m21);
			\vertex [below left = 0.002cm and 0.0015cm of m2] (m22);
			\draw (c2c) -- node[midway, rotate = 53.1301]{$\parallel$} (c2c);
			\draw (v2) -- node[midway]{$\parallel$} (v2);
			\diagram*{
			(f1) -- [very thick, darkred] (f2),
			(c1) -- [photon] (v1),
			(c1c2) -- [fermion] (c1c1),
			(m12) -- [fermion] (m11),
			(m21) -- [fermion] (m22),
			(c2) -- [photon] (v1),
			(v2) -- [half left, fermion] (v1),
			(v2) -- [half right, fermion] (v1),
			};
		\end{feynman}
	\end{tikzpicture}
        \,.
	\label{eq:one-loop_fully-retarded}
\end{align}
In the following, we focus on the bulk part of these diagrams and derive the graviton effective theory after integrating $\chi$ out.
The derived operators can then be contracted with the gravitons on the end-of-inflation hypersurface to obtain the tensor power spectrum.
Anticipating Eq.~\eqref{eq:soft_limit_bulk-to-boundary}, we take the external $\hat{h}_{rij}$ time-independent, and take the spatial momentum $k \to 0$ of the external $\hat{h}_{aij}$ and $\hat{h}_{rij}$.

In the soft limit of the external gravitons, the contribution involving the three-point interactions is given by
\begin{align}
	&\left[\begin{tikzpicture}[baseline=(middle)]
	\begin{feynman}[inline = (base.middle)]
		\vertex (a);
		\vertex [right = 0.65 of a] (v1);
		\vertex [right = 1.25 of v1] (v2);
		\vertex [right = 0.625 of v1] (c);
		\vertex [below = 0.55 of c] (vm);
		\vertex [right = 0.65 of v2] (b);
		\vertex [left = 0.325 of v1] (am);
		\vertex [right = 0.01 of am] (am1);
		\vertex [left = 0.01 of am] (am2);
		\vertex [right = 0.325 of v2] (bm);
		\vertex [right = 0.01 of bm] (bm1);
		\vertex [left = 0.01 of bm] (bm2);
		\begin{pgfonlayer}{bg}
		\draw (vm) -- node[midway]{$\parallel$} (vm);
		\vertex [below = 0.15 of a] (middle);
		\diagram*{
		(vm) -- [in=-90,out=180, fermion] (v1),
		(vm) -- [in=-90,out=0, fermion] (v2) -- [half right, fermion] (v1),
		(v1) -- [photon] (a),
		(v2) -- [photon] (b),
		(bm1) -- [fermion] (bm2);
		(am1) -- [fermion] (am2);
		};
		\end{pgfonlayer}
	\end{feynman}
	\end{tikzpicture}
	\right]_\mathrm{soft}
	\nonumber \\
	&= - \int \dd\tau_1 \dd\tau_2 f^2(\tau_1)f^2(\tau_2) c_s^2(\tau_1) c_s^2(\tau_2)
	\int \frac{\dd^{d-1}l}{(2\pi)^{d-1}}{l}^i{l}^j{l}^k{l}^l
	G_{rr}(l;\tau_1,\tau_2)G_{ra}(l;\tau_1,\tau_2)
	\hat{h}_{aij}(\vec{k},\tau_1) \hat{h}_{rkl}(-\vec{k}),
\end{align}
where $\vec{k}$ is the momentum of the external gravitons (which we take to be soft and keep only the leading order terms), $\vec{l}$ is the momentum of $\chi$ inside the loop, and $G$ denotes the $\chi$ propagators.
Note that we have dropped the time dependence from $\hat{h}_{rij}$,
which is eventually contracted with the graviton on the end-of-inflation hypersurface and gives us the statistical propagator,  since the statistical propagator is time-independent in the soft limit (see Eq.~\eqref{eq:soft_limit_bulk-to-boundary}).
To further simplify this expression, we use a trick and consider the following slightly deformed quadratic action for $\chi$:
\begin{align}
	S_{\chi;\epsilon} = \frac{1}{2}\int \dd^{d-1}x \dd\tau\,f^2(\tau) \left[\left(\frac{\dd\hat{\chi}}{\dd\tau}\right)^2 - c_s^2(\tau)(1+\epsilon)^2(\partial_i \hat{\chi})^2 - m_\chi^2(\tau)\hat{\chi}^2\right].
\end{align}
We can compute the Green's functions of $\chi$ from this action in two ways: (1) including $1+\epsilon$ to all orders, which simply shifts the momentum $k \to (1+\epsilon)k$, and (2) treating $\epsilon$ perturbatively.
Comparing the two results, we obtain diagrammatically
\begin{align}
	\begin{tikzpicture}[baseline=(middle)]
	\begin{feynman}[inline = (base.middle)]
		\vertex (a);
		\vertex [below = 0.095 of a] (middle);
		\vertex [right = 0.65 of a] (v1);
		\vertex [right = 0.65 of v1] (b);
		\begin{pgfonlayer}{bg}
		\draw (v1) -- node[midway]{$\parallel$} (v1);
		\diagram*{
		(v1) -- [fermion, very thick] (b),
		(v1) -- [fermion, very thick] (a),
		};
		\end{pgfonlayer}
	\end{feynman}
	\end{tikzpicture}
	&~=~
	\begin{tikzpicture}[baseline=(middle)]
	\begin{feynman}[inline = (base.middle)]
		\vertex (a);
		\vertex [below = 0.095 of a] (middle);
		\vertex [right = 0.65 of a] (v1);
		\vertex [right = 0.65 of v1] (b);
		\begin{pgfonlayer}{bg}
		\draw (v1) -- node[midway]{$\parallel$} (v1);
		\diagram*{
		(v1) -- [fermion] (b),
		(v1) -- [fermion] (a),
		};
		\end{pgfonlayer}
	\end{feynman}
	\end{tikzpicture}
	~+~
	\begin{tikzpicture}[baseline=(middle)]
	\begin{feynman}[inline = (base.middle)]
		\vertex (a);
		\vertex [below = 0.095 of a] (middle);
		\vertex [right = 0.65 of a] (v1);
		\vertex [right = 0.65 of v1] (v2);
		\vertex [right = of v2] (b);
		\begin{pgfonlayer}{bg}
		\draw (v1) -- node[midway]{$\parallel$} (v1);
		\draw (v2) -- node[midway]{$\times$} (v2);
		\diagram*{
		(v1) -- [fermion] (v2) -- [fermion] (b),
		(v1) -- [fermion] (a),
		};
		\end{pgfonlayer}
	\end{feynman}
	\end{tikzpicture}
	~+ 
	\begin{tikzpicture}[baseline=(middle)]
	\begin{feynman}[inline = (base.middle)]
		\vertex (a);
		\vertex [below = 0.095 of a] (middle);
		\vertex [right = of a] (v1);
		\vertex [right = 0.65 of v1] (v2);
		\vertex [right = 0.65 of v2] (b);
		\draw (v1) -- node[midway]{$\times$} (v1);
		\begin{pgfonlayer}{bg}
		\draw (v2) -- node[midway]{$\parallel$} (v2);
		\diagram*{
		(v2) -- [fermion] (v1) -- [fermion] (a),
		(v2) -- [fermion] (b),
		};
		\end{pgfonlayer}
	\end{feynman}
	\end{tikzpicture}
	~+\cdots,
\end{align}
where the thick line on the left-hand side indicates the propagator with $\epsilon$ included to all orders,
the cross, ``$\times$,'' represents the operator $\epsilon\nabla^2$ insertion,
and the dots indicate higher order terms in $\epsilon$.
Analytically this is expressed as
\begin{align}
	G_{rr}((1+\epsilon)l;&\,\tau_1,\tau_2) - G_{rr}(l;\tau_1,\tau_2) 
	\nonumber \\
	&= -2i \epsilon l^2\int \dd\tau f^2(\tau) c_s^2(\tau) \left[G_{rr}(l;\tau_1,\tau)G_{ra}(l;\tau_2,\tau) + G_{ra}(l;\tau_1,\tau)G_{rr}(l;\tau_2,\tau)\right]
	+ \cdots.
\end{align}
By expanding the left-hand side with respect to $\epsilon$ and comparing the linear terms in $\epsilon$, we obtain\footnote{
    The right-hand side contains a folded singularity as $\tau \to -\infty$ unless the initial condition is taken 
    as the Bunch-Davies vacuum~\cite{Flauger:2013hra,Arkani-Hamed:2015bza,Arkani-Hamed:2018kmz,Green:2020whw}.
    As we have mentioned, in this paper, we focus on the Bunch-Davies vacuum.
}
\begin{align}
	\frac{\partial}{\partial \log l}G_{rr}(l;\tau_1,\tau_2)
	=  -2i l^2\int \dd\tau f^2(\tau) c_s^2(\tau) \left[G_{rr}(l;\tau_1,\tau)G_{ra}(l;\tau_2,\tau) + G_{ra}(l;\tau_1,\tau)G_{rr}(l;\tau_2,\tau)\right].
	\label{eq:Grr_logl_deriv}
\end{align}
As clear from the derivation, this relation holds for arbitrary $f(\tau)$, $c_s(\tau)$, and $m_\chi(\tau)$ as long as one derives the Green's functions by solving the mode equation, which involves $f(\tau)$, $c_s(\tau)$, and $m_\chi(\tau)$ consistently.
It is easy to check this relation explicitly in simple cases, such as a massless scalar field in de~Sitter spacetime where $f = a = -1/H\tau$, $c_s = 1$, and $m_\chi = 0$.
By noting that $G_{rr}(l;\tau_1,\tau_2) = G_{rr}(l;\tau_2,\tau_1)$ and using Eq.~\eqref{eq:Grr_logl_deriv}, we obtain
\begin{align}
	\left[\begin{tikzpicture}[baseline=(middle)]
	\begin{feynman}[inline = (base.middle)]
		\vertex (a);
		\vertex [right = 0.65 of a] (v1);
		\vertex [right = 1.25 of v1] (v2);
		\vertex [right = 0.625 of v1] (c);
		\vertex [below = 0.55 of c] (vm);
		\vertex [right = 0.65 of v2] (b);
		\vertex [left = 0.325 of v1] (am);
		\vertex [right = 0.01 of am] (am1);
		\vertex [left = 0.01 of am] (am2);
		\vertex [right = 0.325 of v2] (bm);
		\vertex [right = 0.01 of bm] (bm1);
		\vertex [left = 0.01 of bm] (bm2);
		\begin{pgfonlayer}{bg}
		\draw (vm) -- node[midway]{$\parallel$} (vm);
		\vertex [below = 0.15 of a] (middle);
		\diagram*{
		(vm) -- [in=-90,out=180, fermion] (v1),
		(vm) -- [in=-90,out=0, fermion] (v2) -- [half right, fermion] (v1),
		(v1) -- [photon] (a),
		(v2) -- [photon] (b),
		(bm1) -- [fermion] (bm2);
		(am1) -- [fermion] (am2);
		};
		\end{pgfonlayer}
	\end{feynman}
	\end{tikzpicture}
	\right]_\mathrm{soft}
	&= -\frac{i}{4} \int \dd\tau f^2(\tau) c_s^2(\tau) \int\frac{\dd^{d-1}l}{(2\pi)^{d-1}}
	\frac{{l}^i{l}^j{l}^k{l}^l}{l^2} \left[\frac{\partial}{\partial\log l}G_{rr}(l;\tau,\tau)\right] \hat{h}_{aij}(\vec{k},\tau) \hat{h}_{rkl}(-\vec{k}).
\end{align}
It is crucial here that $\hat{h}_{rij}$ does not depend on time and the $\chi$ propagators depend only on $\vec{l}$ and not on $\vec{k}$.
This simplification thus holds only in the soft limit.
Since Green's function depends only on the size of $\vec{l}$, or $l$, the angular integral gives
\begin{align}
	\int\frac{\dd^{d-1}l}{(2\pi)^{d-1}}\frac{{l}^i{l}^j{l}^k{l}^l}{l^2}
	= \frac{1}{(d-1)(d+1)}\left(\delta^{ij}\delta^{kl} + \delta^{ik}\delta^{jl} + \delta^{il}\delta^{jk} \right)
	\int\frac{\dd^{d-1}l}{(2\pi)^{d-1}}l^2,
\end{align}
and hence, after integration by parts,\footnote{
	The IR contribution vanishes for $d = 4$ and $\lim_{l\to 0} G_{rr} \propto l^{-3}$, which is the case of our interest,
    while the UV contribution should be dealt with by a regularization that does not break the symmetry of the original theory,
    such as the dimensional regularization.
	These contributions from the integral boundaries are anyway irrelevant for us since the enhanced small-scale scalar perturbations of our interest
	are with finite momenta, and not in the limit $l \to 0$ nor $l \to \infty$.
}
we obtain
\begin{align}
	\left[
	\begin{tikzpicture}[baseline=(middle)]
	\begin{feynman}[inline = (base.middle)]
		\vertex (a);
		\vertex [right = 0.65 of a] (v1);
		\vertex [right = 1.25 of v1] (v2);
		\vertex [right = 0.625 of v1] (c);
		\vertex [below = 0.55 of c] (vm);
		\vertex [right = 0.65 of v2] (b);
		\vertex [left = 0.325 of v1] (am);
		\vertex [right = 0.01 of am] (am1);
		\vertex [left = 0.01 of am] (am2);
		\vertex [right = 0.325 of v2] (bm);
		\vertex [right = 0.01 of bm] (bm1);
		\vertex [left = 0.01 of bm] (bm2);
		\begin{pgfonlayer}{bg}
		\draw (vm) -- node[midway]{$\parallel$} (vm);
		\vertex [below = 0.15 of a] (middle);
		\diagram*{
		(vm) -- [in=-90,out=180, fermion] (v1),
		(vm) -- [in=-90,out=0, fermion] (v2) -- [half right, fermion] (v1),
		(v1) -- [photon] (a),
		(v2) -- [photon] (b),
		(bm1) -- [fermion] (bm2),
		(am1) -- [fermion] (am2),
		};
		\end{pgfonlayer}
	\end{feynman}
	\end{tikzpicture}
	\right]_\mathrm{soft}
	&= \frac{i}{2(d-1)}\int \dd\tau f^2(\tau) c_s^2(\tau) \left[\int \frac{\dd^{d-1}l}{(2\pi)^{d-1}} l^2 G_{rr}(l;\tau,\tau)\right]\hat{h}_{a}^{ij}(\vec{k},\tau) 
	\hat{h}_{rij}(-\vec{k}),
\end{align}
where we used that $h_{ij}$ is traceless.
The contribution from the four-point interaction is easily computed and is given in the soft limit as
\begin{align}
	\left[
	\begin{tikzpicture}[baseline=(middle)]
	\begin{feynman}[inline = (base.middle)]
		\vertex (a);
		\vertex [right = of a] (v1);
		\vertex [below = of v1] (v2);
		\vertex [right = of v1] (b);
		\begin{pgfonlayer}{bg}
		\draw (v2) -- node[midway]{$\parallel$} (v2);
		\vertex [below = 0.6 of a] (middle);
		\vertex [left = 0.6 of v1] (am);
		\vertex [right = 0.01 of am] (am1);
		\vertex [left = 0.01 of am] (am2);
		\vertex [right = 0.6 of v1] (bm);
		\vertex [right = 0.01 of bm] (bm1);
		\vertex [left = 0.01 of bm] (bm2);
		\diagram*{
		(v1) -- [half left, anti fermion] (v2),
		(v1) -- [half right, anti fermion] (v2),
		(v1) -- [photon] (b),
		(v1) -- [photon] (a),
		(bm1) -- [fermion] (bm2),
		(am1) -- [fermion] (am2),
		};
		\end{pgfonlayer}
	\end{feynman}
	\end{tikzpicture}
	\right]_\mathrm{soft}
	&=  -\frac{i}{2(d-1)}\int \dd\tau f^2(\tau) c_s^2(\tau) \left[\int \frac{\dd^{d-1}l}{(2\pi)^{d-1}} l^2 G_{rr}(l;\tau,\tau)\right]\hat{h}_{a}^{ij}(\vec{k},\tau) 
	\hat{h}_{rij}(-\vec{k}),
\end{align}
where we have used the angular integral identity,
\begin{align}
	\int \frac{\dd^{d-1}l}{(2\pi)^{d-1}} l^i l^j G_{rr}(l;\tau,\tau)
	&= \frac{\delta^{ij}}{d-1}\int \frac{\dd^{d-1}l}{(2\pi)^{d-1}} l^2 G_{rr}(l;\tau,\tau).
\end{align}
By combining the two contributions, we conclude that
\begin{align}
	\left[
	\begin{tikzpicture}[baseline=(middle)]
	\begin{feynman}[inline = (base.middle)]
		\vertex (a);
		\vertex [right = 0.65 of a] (v1);
		\vertex [right = 1.25 of v1] (v2);
		\vertex [right = 0.625 of v1] (c);
		\vertex [below = 0.55 of c] (vm);
		\vertex [right = 0.65 of v2] (b);
		\vertex [left = 0.325 of v1] (am);
		\vertex [right = 0.01 of am] (am1);
		\vertex [left = 0.01 of am] (am2);
		\vertex [right = 0.325 of v2] (bm);
		\vertex [right = 0.01 of bm] (bm1);
		\vertex [left = 0.01 of bm] (bm2);
		\begin{pgfonlayer}{bg}
		\draw (vm) -- node[midway]{$\parallel$} (vm);
		\vertex [below = 0.15 of a] (middle);
		\diagram*{
		(vm) -- [in=-90,out=180, fermion] (v1),
		(vm) -- [in=-90,out=0, fermion] (v2) -- [half right, fermion] (v1),
		(v1) -- [photon] (a),
		(v2) -- [photon] (b),
		(bm1) -- [fermion] (bm2),
		(am1) -- [fermion] (am2),
		};
		\end{pgfonlayer}
	\end{feynman}
	\end{tikzpicture}
	\,+\,
	\begin{tikzpicture}[baseline=(middle)]
	\begin{feynman}[inline = (base.middle)]
		\vertex (a);
		\vertex [right = of a] (v1);
		\vertex [below = of v1] (v2);
		\vertex [right = of v1] (b);
		\begin{pgfonlayer}{bg}
		\draw (v2) -- node[midway]{$\parallel$} (v2);
		\vertex [below = 0.6 of a] (middle);
		\vertex [left = 0.6 of v1] (am);
		\vertex [right = 0.01 of am] (am1);
		\vertex [left = 0.01 of am] (am2);
		\vertex [right = 0.6 of v1] (bm);
		\vertex [right = 0.01 of bm] (bm1);
		\vertex [left = 0.01 of bm] (bm2);
		\diagram*{
		(v1) -- [half left, anti fermion] (v2),
		(v1) -- [half right, anti fermion] (v2),
		(v1) -- [photon] (b),
		(v1) -- [photon] (a),
		(bm1) -- [fermion] (bm2),
		(am1) -- [fermion] (am2),
		};
		\end{pgfonlayer}
	\end{feynman}
	\end{tikzpicture}
	\right]_\mathrm{soft}
	&~= 0,
\end{align}
in the soft limit of the external gravitons.
In other words, the one-loop correction to the tensor power spectrum vanishes in the superhorizon limit,
and there is no scale-invariant correction to the tensor power spectrum from the small-scale (potentially enhanced) scalar perturbations.

In our proof of the cancellation, the identity~\eqref{eq:Grr_logl_deriv} plays the central role.
It involves one Green's function on the left-hand side and two Green's functions on the right-hand side, and therefore relates diagrams with distinct topologies, \emph{i.e.}, with different numbers of propagators.
Indeed, our calculation shows that the first diagram in Eq.~\eqref{eq:one-loop_fully-retarded}, after using Eq.~\eqref{eq:Grr_logl_deriv}, 
reduces to the same topology as the second one in the soft limit, and the relative coefficients are such that the sum cancels out.
The cancellation of diagrams with distinct topologies in a specific limit 
is a recurrent feature of theories with non-trivial symmetries and hence Ward-Takahashi identities, such as the gauge theory
(see \textit{e.g.}, Refs.~\cite{Sterman:1993hfp,Veltman:1994wz} and Refs.~\cite{Creminelli:2012ed,Hinterbichler:2012nm,Garriga:2013rpa,Hinterbichler:2013dpa,Berezhiani:2013ewa,Tanaka:2015aza,Tanaka:2017nff,Arkani-Hamed:2018kmz,Hui:2018cag}).
It is therefore natural to expect that our calculation can be understood from the viewpoint of the Ward-Takahashi identity, and we will come back to this point in a separate publication~\cite{our2nd}.

\subsubsection*{Comment on the literature}

The one-loop correction of an enhanced (spectator) scalar field to the tensor power spectrum was computed in the past.
In particular, Refs.~\cite{Ota:2022hvh,Ota:2022xni} claim that there is a scale-invariant correction to the tensor perturbation from the spectator scalar field, even in the soft limit, in direct contradiction with our result.
In the following, we explain (what we think is) the origin of the discrepancy.

As emphasized, Eq.~\eqref{eq:Grr_logl_deriv} is of central importance for the cancellation of the one-loop correction in the soft limit.
This identity holds for an arbitrary background time-dependence of $f(\tau)$, $c_s(\tau)$, and $m_\chi (\tau)$ as long as one properly solves the mode equation, in the presence of $f(\tau)$, $c_s(\tau)$, and $m_\chi (\tau)$, to derive Green's functions.
On the other hand, Refs.~\cite{Ota:2022hvh,Ota:2022xni} ``modeled'' the enhancement of the small-scale scalar perturbations by modifying the mode functions artificially, without explicitly solving the mode equations.
That is to say, the enhancement of the small-scale perturbations is inserted by hand, which is inconsistent with their assumption of $f(\tau) = a(\tau)$, $c_s (\tau) = 1$, and $m_\chi (\tau) = 0$.
Therefore, their background evolution and the mode functions are not compatible with each other.
Indeed, we have checked that their Eq.~(16) in Ref.~\cite{Ota:2022hvh}, together with the standard evolution of the scale factor $a = -1/H\tau$ during inflation (which we assume to be used in their calculation), does not satisfy Eq.~\eqref{eq:Grr_logl_deriv}.
It is then no surprise that they obtained an unphysical result.
Our result tells us that one should not modify the mode functions arbitrarily without changing the time evolution of $f(\tau)$, $c_s(\tau)$, or $m_\chi(\tau)$ accordingly, and vice versa.

\subsection{Symmetry of soft tensor EFT}
\label{subsec:symmetry}

Before closing this section, we briefly comment on the symmetry of the soft tensor
on the Schwinger--Keldysh contour, which is obtained after integrating out the small-scale perturbations; in our case, the $\chi$ field.
After taking the $\zeta$-gauge, the residual gauge freedom is given by the infinitesimal spatial diffeomorphism generated by $\xi^i$.
Let us consider the following infinitesimal transformation: $\xi_i = M_{ij} x^j$ with $M_{ij}$ being a constant symmetric traceless matrix.
The soft tensor perturbation $h_{ij}$ transforms as
\begin{equation}
	\delta h_{ij} = 2 M_{ij}, \label{eq:linear}
\end{equation}
at a linear order in $\xi_i$.

In the in-in formalism, the soft tensor perturbation is doubled, \textit{i.e.,} $h_{ij}^+$ and $h_{ij}^-$, due to the Schwinger--Keldysh contour (Fig.~\ref{fig:in-in_contour}).
While $h_{ij}^+$ and $h_{ij}^-$ are defined on the different complex time contours, both of them live in the same space for the spatial coordinates.
Hence, the same spatial diffeomorphism denoted above is applied to both $h_{ij}^+$ and $h_{ij}^-$, which implies
\begin{equation}
	\delta h_{aij} = \delta h^+_{ij} - \delta h^-_{ij} = 0, \qquad
	\delta h_{rij} = \frac{1}{2}\left( \delta h^+_{ij} + \delta h^-_{ij} \right) = 2 M_{ij}.
\end{equation}
The soft tensor EFT must be consistent with this symmetry.

Now let us identify the terms in the soft tensor EFT that are consistent with this symmetry at the lowest order in the gradient expansion.
Unitarity forbids those terms solely composed of the $r$-fields, $h_{rij}$ (see Ref.~\cite{Ema:2024hkj}).
The lowest-order term linear in $h_{aij}$ is $h_{a}^{ij} h_{rij}$.
However, this term is not consistent with the symmetry above because $\delta (h_{a}^{ij} h_{rij}) = h_a^{ij} \delta h_{rij} = 2 h_a^{ij} M_{ij}$ is non-vanishing.
This is the intrinsic reason why the term $h_a^{ij} h_{rij}$ vanishes in our explicit calculation.\footnote{
	The same conclusion follows from the symmetry-oriented arguments  
	based on soft de~Sitter effective theory~\cite{Cohen:2020php,Green:2024fsz}.
	Our approach here is based on the Schwinger--Keldysh effective field theory, which was recently developed to fill in the gap between macroscopic non-equilibrium description and microscopic quantum field theory, such as the derivation of the diffusion equation, hydrodynamics, and so on (see Ref.~\cite{Liu:2018kfw} and therein).
	The Schwinger--Keldysh effective field theory is based on unitarity and symmetries of the underlying microscopic theory, and is not limited to de~Sitter spacetime, and thereby the proof of the vanishing $h_a^{ij} h_{rij}$ does not require any specific assumptions about the background geometry.
}
On the other hand, the other quadratic term $h_a^{ij} h_{aij}$ is allowed by the symmetry.
Indeed, this is expected from the fact that the induced gravitational wave from the second-order scalar perturbations originates from 
the (non-local) $h_a^{ij} h_{aij}$ operator (see \textit{e.g.}, Ref.~\cite{Tada:2023rgp}).
As explained already below Eq.~\eqref{eq:soft_limit_bulk-to-boundary}, the power spectrum originated from $h_a^{ij} h_{aij}$ obeys the causality suppression of $k^3$ in the soft limit.

\section{Conclusion}
\label{sec:conclusion}

In this paper, we have considered loop effects of scalar fluctuations on the tensor power spectrum by integrating out a scalar field in a general background over the Schwinger--Keldysh contour to derive a soft EFT for the tensor mode.
We have explicitly shown that the resulting soft EFT for the tensor never contains terms that lead to scale-invariant corrections.
This integration is most easily performed by the Schwinger--Keldysh $r/a$ basis, and indeed there is no term
of the form $h_{aij}h^{ij}_r$ that, if present, would lead to scale-invariant corrections.
We have also demonstrated that the absence of this term is guaranteed by the diffeomorphism invariance of GR, while the term $h_{aij}h^{ij}_a$ allowed by the symmetry is consistent with the causality suppression of $k^3$ in the soft limit.
We have confirmed the cancellation explicitly at the one-loop level,
while our argument based on symmetry of soft EFT holds to all-loop order.
\footnote{
	We emphasize again that we have focused on the leading-order term in the soft limit at the level of the conformal time integrand,
	which are potentially sensitive to the dynamics at the intermediate time scale such as the transient USR phase.
	Even though the sub-leading terms at the integral level can generate non-zero scale-invariant corrections after the conformal time integral, 
	these terms are sensitive only to the dynamics in the infinite past, $\tau \sim -1/k \to -\infty$, and are out of our interest 
	(see the discussion at the end of Sec.~\ref{sec:ra_basis}).
}

In our proof, the identity involving Green's functions, Eq.~\eqref{eq:Grr_logl_deriv}, plays an essential role for the cancellation.
We emphasize that this relation holds for an arbitrary background time evolution.
Therefore, if one explicitly solves the mode equations in a given background time evolution, such as the transient ultra-slow-roll phase, to calculate loop corrections, it would be a useful consistency check to see if the derived mode functions indeed satisfy Eq.~\eqref{eq:Grr_logl_deriv}.

Several extensions of our analysis are on the waiting list.
As anticipated by our symmetry arguments, similar non-trivial cancellations are expected for multi-loop corrections, and corrections to higher-point correlators of soft tensor modes from the small-scale perturbations.
In the proof by explicit calculations, the cancellation among diagrams of different topologies plays a crucial role, whose relation to the Ward--Takahashi identity is yet to be understood.\footnote{
	See Ref.~\cite{Tanaka:2015aza} for the study on the relation between the Ward-Takahashi idenitity and the conservation of the superhorizon curvature perturbations.
}
Our analysis is not limited to the spectator field $\chi$ but can be extended to the case where the small-scale perturbations are sourced by the inflaton field.
Moreover, the idea of the soft EFT should be applicable to the scalar perturbations as well, and the analysis of the soft scalar EFT is in progress.
All these analyses will be reported in a separate publication~\cite{our2nd}.

\paragraph{Acknowledgments}

The authors are grateful to Keisuke Inomata, Yuichiro Tada, Takahiro Terada and Jun'ichi Yokoyama for useful discussions.
M.\,H.~is supported by Grant-in-Aid for JSPS Fellows 23KJ0697.
R.\,J.~is supported by JSPS KAKENHI Grant Numbers 23K17687 and 24K07013.
K.\,M.~is supported by JSPS KAKENHI Grant No.~JP22K14044.

\small
\bibliographystyle{utphys}
\bibliography{ref}

\end{document}